\documentclass[journal]{IEEEtran}
\IEEEoverridecommandlockouts
\usepackage{amsmath,amssymb,amsfonts}
\usepackage{graphicx}
\usepackage{textcomp}
\usepackage{xcolor}
\usepackage{bm}
\usepackage{verbatim}
\usepackage{algorithm} 
\usepackage{algpseudocode} 
\usepackage{cite}
\usepackage{stfloats}

\newtheorem{theorem}{\textbf{Theorem}}
\newtheorem{definition}{Definition}
\newtheorem{lemma}{\textbf{Lemma}}

\ifCLASSINFOpdf
\else
\fi
\begin{document}
\title{Joint Optimization of Sensing and Computation for Status Update in Mobile Edge Computing Systems}
	
	\author{Yi Chen, Zheng Chang,~\IEEEmembership{Senior Member, IEEE,}   Geyong~Min,~\IEEEmembership{Senior Member,~IEEE,}     Shiwen~Mao,~\IEEEmembership{Fellow,~IEEE,} and T. H\"am\"al\"ainen,~\IEEEmembership{Senior Member,~IEEE}
 		
\thanks{Y. Chen and Z. Chang are with School of Computer Science and Engineering, University of Electronic Science and Technology of China, 611731 Chengdu, China. G. Min is with Department of Computer Science, University of Exeter, Exeter, EX4 4QF, U.K.. S. Mao is with Department of Electrical and Computer Engineering, Auburn University, AL, USA. T. H\"am\"al\"ainen is with Faculty of Information Technology, University of Jyv\"askyl\"a, P. O. Box 35, FIN-40014 Jyv\"askyl\"a, Finland.}
\thanks{Manuscript received April 19, 2005, revised August 26, 2015.}}

\markboth{Journal of \LaTeX\ Class Files,~Vol.~14, No.~8, August~2015}%
{Shell \MakeLowercase{\textit{et al.}}: Bare Demo of IEEEtran.cls for IEEE Journals}
\maketitle

\begin{abstract}
IoT devices recently are utilized to detect the state transition in the surrounding environment and then transmit the status updates to the base station for future system operations. To satisfy the stringent timeliness requirement of the status updates for the accurate system control, age of information (AoI) is introduced to quantify the freshness of the sensory data. Due to the limited computing resources, the status update can be offloaded to the mobile edge computing (MEC) server for execution to ensure the information freshness. Since the status updates generated by insufficient sensing operations may be invalid and cause additional processing time, the data sensing and processing operations need to be considered simultaneously. In this work, we formulate the joint data sensing and processing optimization problem to ensure the freshness of the status updates and reduce the energy consumption of IoT devices. Then, the formulated NP-hard problem is decomposed into the sampling, sensing and computation offloading optimization problems. Afterwards, we propose a multi-variable iterative system cost minimization algorithm to optimize the system overhead. Simulation results show the efficiency of our method in decreasing the system cost and dominance of sensing and processing under different scenarios. 
\end{abstract}

\begin{IEEEkeywords}
Age of information, mobile edge computing, computation offloading, status update.
\end{IEEEkeywords}
\IEEEpeerreviewmaketitle

\bibliographystyle{IEEEtran}
\section{Introduction}
\subsection{Background and Motivation}
With the development of Internet of Things (IoT) Infrastructure, ubiquitous connection of billions of IoT devices ranging from tiny IoT sensors to the more powerful smarts phones is enabled to be realistic~\cite{xuoptimal}. To emerge the various IoT applications like object recognition, traffic monitoring and autonomous driving, vast information from the physical world should be extracted and transformed into status updates to realize the intelligent control for IoT devices~\cite{pengsense}. Considering the processing operations of the status update are commonly time-consuming and computation-intensive, it is a tough task to process the sensory data in a timely manner for IoT devices with limited storage and computation capacity. Mobile edge computing (MEC) has been regarded as a promising technology to overcome the resource constraints of IoT devices by yielding cloud-like computing resources~\cite{zhangjoint,elegendyefficent}. In this case, IoT devices are able to offload the computing tasks to the nearby MEC server for further execution. Through deploying powerful computing resources in proximity to IoT devices and executing computing tasks on behalf of IoT devices, the computing and storage pressure of IoT devices can be relieved~\cite{maomec,ningdynamic}.\par

Moreover, the accurate monitoring of the IoT system has a strict requirement on the freshness of the collected information. Recently, age of information (AoI) which is defined as the time elapsed since the generation of the last status update is adopted as a performance metric to quantify the freshness of generated status information~\cite{Kaulmini}. When an IoT device generates and receives a status update successfully, its value of AoI is reset to 0. The AoI of the status update increases linearly with time until the next status update is successfully received. The average value of the AoI during the continuous sensing periods reflects the freshness of the sensory data~\cite{sangpower}. By introducing the concept of AoI, the abstract information freshness problem can be transformed into a concrete mathematical optimization problem. \par

The typical IoT system can be constructed as a three-layered structure including the sensing layer, the network layer and the application layer~\cite{atzoriiot}. In the sensing layer, the IoT devices will keep sensing the state transition process and generate state updates when deemed necessary. Then, the sensory data are transmitted through the network layer for processing and future system control. By repeatedly receiving the valid status update from different IoT devices, the AoI at the BS can be reduced during the system control process~\cite{zhangsense}. To clarify our work in this paper, sensing and processing are explicitly defined in this paper as follows. 
 
 \begin{definition}   
Sensing of the IoT device is defined as the process that the IoT device observes the environment transition and generates the status update when necessary.
  \end{definition}
    
\begin{definition}
Processing is defined as determining whether a status update is valid and extracting the information from the status update necessary to perform system control.
  \end{definition}

\begin{figure}[t]
    \centering
    \includegraphics[width=8cm]{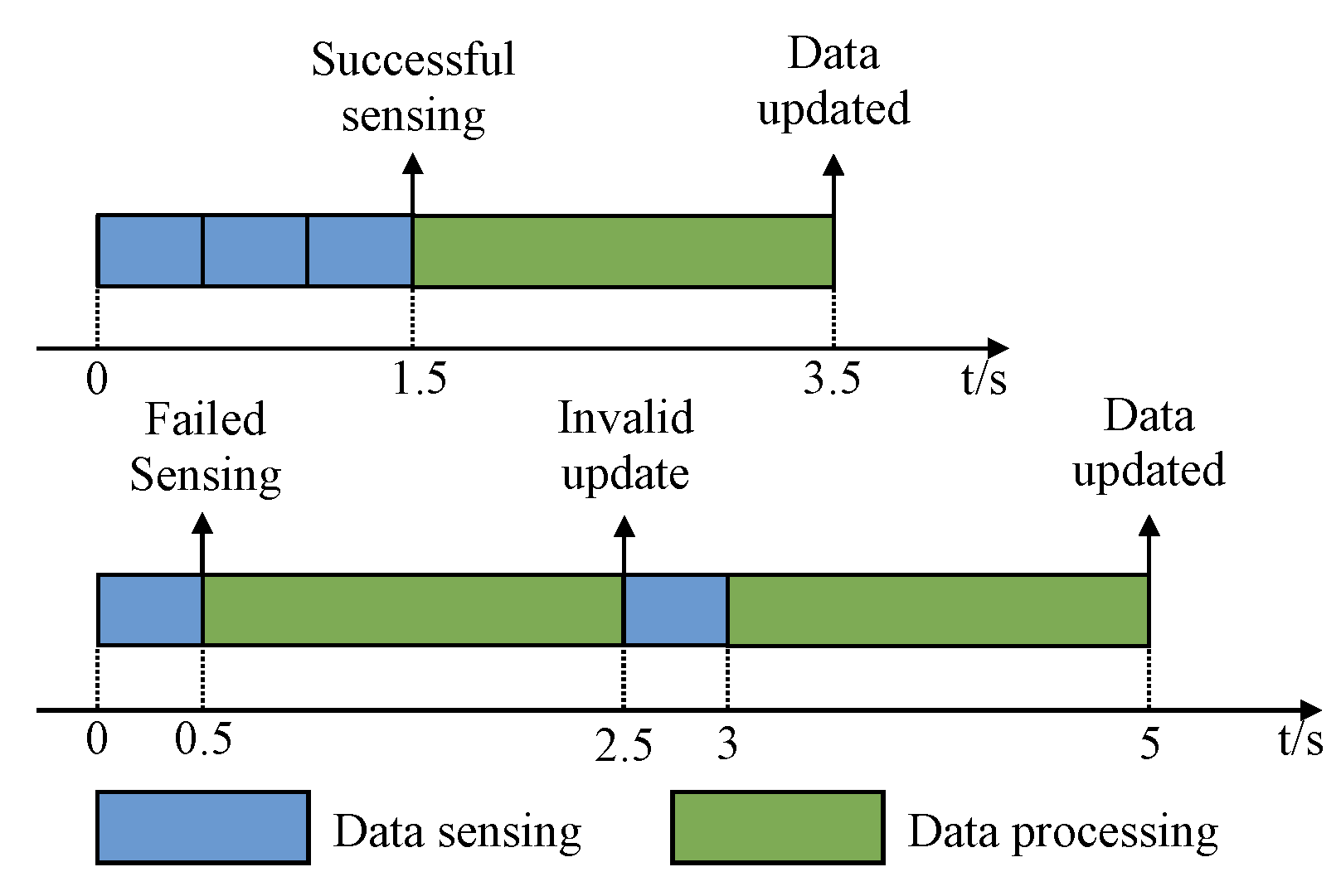}
    \caption{Example of the sensing-processing procedure.}
    \label{figexp}
\end{figure}

However, in the sensing layer, in order to ensure the information freshness on the BS side, IoT devices should perform the sampling operations and generate status updates at as high a frequency as possible. With the short sampling interval, the BS can achieve a low value of AoI due to the frequent status updates~\cite{liaop}. But sampling the state transition frequently brings additional energy consumption, which is not negligible for IoT devices with limited battery capacity. Moreover, successfully sensing the state updates may be a random event for IoT devices, due to the possible state tracking error rate~\cite{kamtowards,ayanaoi}. Longer sensing time can significantly improve the sensing successful rate, at the cost of additional time overhead and freshness of status updates. In addition, if the sensing time is insufficient, multiple sensing failures can lead to multiple unnecessary repetitions of sensing process. As the data sensing-processing procedure shown in Fig. \ref{figexp}, the IoT device that performs sensing operations three times will receive a longer sensing time and a short total task duration time. Nevertheless, the IoT device only performs a single sensing operation generates the invalid update unfortunately and has to execute the extra sensing-processing procedure until the sensing operation is successful. Thus, the sensing time of the IoT device should be considered to obtain the minimum time overhead. 

In the processing part, since some information embedded in the sensory data requires further processing before it can be utilized for further system control, the execution time of the sensing task of the IoT device also has a significant impact on the freshness of the status update. Due to the computation capacity limitations of the IoT devices themselves, they would prefer to offload the tasks to the MEC server for execution to obtain shorter processing time. But selfish IoT devices will compete for limited computational and communication resources, which may create utility conflicts and increase the system overhead during the computation offloading decision-making process~\cite{wangcomputation,zhaocomputation}.
\subsection{Contribution}

In view of these challenges mentioned above, we further analyze the joint optimization problem of sensing and processing in MEC system. By investigating the effects of sampling, sensing, and processing on system overhead independently, we seek to realize the IoT device sensing and processing trade-off in terms of the information freshness and energy consumption for IoT devices. The main contributions of the paper can be summarized as follows.
\begin{itemize}
    \item We first introduce a MEC-assisted IoT system, where the IoT devices keep generating status update periodically. Then, the state sensing, data transmission and task processing are modeled separately and the AoI of the sensory data during the status updating process is formulated. 
    \item We formulate a joint sensing and processing optimization problem for status updates to minimize the system overhead including the energy consumption and the information freshness. Then, the NP-hard problem is transformed into three subproblems to optimize the sampling interval, sensing time and computation offloading decision individually.
    
    \item We solve the sensing and sampling subproblems with extremum principles, and solve the computation offloading decision-making problem with a game-theoretic approach. Then, a multi-variable iteration system cost optimization algorithm (MISCO) is proposed to minimize the system overhead.
    
    \item We conduct the simulation experiments to prove the effectiveness of our proposed optimization algorithm. Numerical results illustrates the connection between the data sensing and processing. 
 \end{itemize}
 \subsection{Organization}

The rest of this paper is organized as follows. In Section \ref{rw}, we discuss the related works. In Section \ref{model}, we present the MEC-assist IoT system model and formulate the sensing, transmission and processing models during the status update process. In Section \ref{AoI}, we analyze the AoI in the considered system. In Section \ref{pf}, we propose the system overhead optimization problem and decompose the problem into sensing, sampling and processing subproblems. In Section \ref{method}, the joint sensing and processing optimization algorithm MISCO is proposed. In Section \ref{expresult}, we show the numerical results of the simulation experiment. Finally in Section \ref{conclusion}, we conclude our paper.

\section{Related Work}\label{rw}
Freshness of the status update has emerged as a recent highlight in the field of network research, which leads to the increasing research interest in AoI served as the metrics to measure the freshness of information \cite{kaulaoi2019,kaulaoi2021,kadotaoptimize,fengaoi,zhoujoint,chenaoi}. Yates \textit{et al.} in~\cite{kaulaoi2019} investigate real-time status updates generated by multiple independent sources sending to a single monitor with an AoI timeliness metric and derive the general values of AoI suitable for various multi-source service system. Kadota \textit{et al.} in~\cite{kadotaoptimize} study a single-hop wireless network where multiple nodes transmits time-sensitive information to the base station while minimizing the expected weighted total AoI of the network and satisfying the just-in-time throughput at the same time. Feng \textit{et al.} in~\cite{fengaoi} design an optimal strategy for the energy harvesting sensor to generate status updates with the purpose of minimizing the long-term average AoI and satisfying the energy constraint in the different cases of whether the system has updating feedback. Zhou \textit{et al.} in~\cite{zhoujoint} study a time-intensive IoT monitoring system where IoT devices continuous generate and transmit the status updates with updating cost. Through simultaneously optimizing the sampling and updating process, the minimum average AoI of the destination node is derived under the upper bound of the updating cost. Chen \textit{et al.} in~\cite{chenaoi} investigate the AoI-aware radio resource management problem in a Manhattan Grid vehicle-to-vehicle network to realize the optimal frequency allocation and packet scheduling decision-making.  

Several relative works have been conducted in the context of the optimization of data sensing~\cite{pengsense,zhangsense,liuuav,zhaosecure,hucooperative}. Peng \textit{et al.} in~\cite{pengsense} propose a joint sensing and communication scheduling framework for status update in the multi-access network to minimize the average status error. The trade-off between the data sensing and data transmission are investigated to minimize the long-term average AoI during multiple sensing cycles in~\cite{zhangsense,zhaosecure}. In~\cite{liuuav,hucooperative}, UAV trajectory optimization problem where UAV performs the sensing tasks to collect the time-intensive data are studied to satisfy the system AoI threshold. 

In addition to the issues mentioned above, the collected data should be offloaded to the MEC server for low-latency processing. Due to the resource competition among IoT devices, computation offloading optimization problem need to be considered. Computation offloading decision-making problem has been widely investigated to address the computation capacity constraint and communication interference. Game-theoretic method has been introduced to address the computation offloading decision-making problem~\cite{chendecentralized,yangd2d,dingpotential,wanggame}. Chen \textit{et al.} in~\cite{chendecentralized} first utilize the concept of the potential game to achieve the Nash equilibrium of the computation offloading game. Yang \textit{et al.} in ~\cite{yangd2d} propose a computation offloading game including multiple computation offloading schemes which take advantage of the available resources of the idle mobile devices.

However, most of the studies have focused on reducing processing latency and system energy consumption while performing computation offloading decision making, ignoring the role of information freshness in improving service quality. In addition, the state-sensing procedure may generate invalid state updates, resulting in extra processing time and deteriorating the service quality. Therefore, the sensing and processing operations need to be considered jointly to ensure the freshness of the status update and minimize the system overhead.
\section{System Model}\label{model}
\subsection{Network Model}
\begin{figure}[t]
    \centering
    \includegraphics[width=8cm]{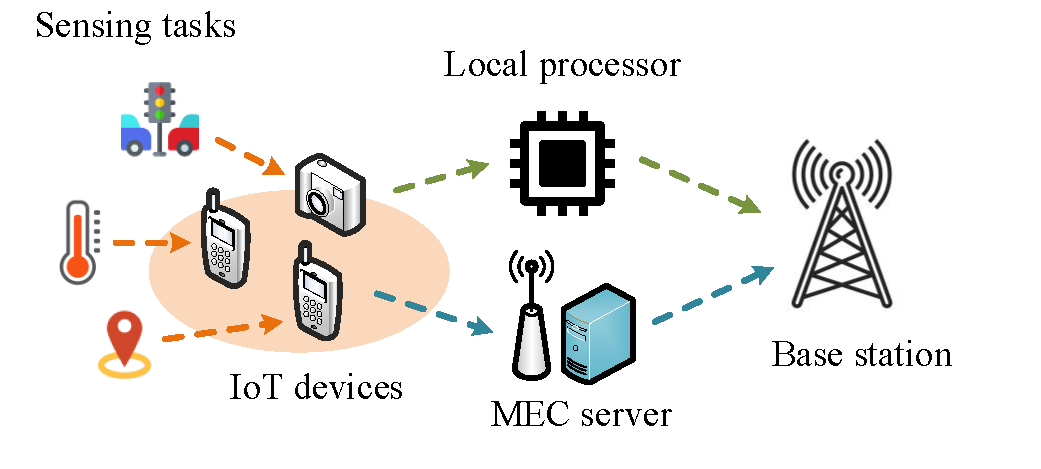}
    \caption{Example of MEC-assisted IoT system.}
    \label{fig1}
\end{figure}

We consider a typical MEC-assisted IoT system with a set $\mathcal{N}$ of $N$ IoT devices and an MEC server. The IoT devices monitor a physical process like the traffic condition in the autonomous driving system. When deemed necessary, the IoT devices sample the status information and generate the status update which need to be processed further to achieve the accurate control. The status sampling process of each IoT device $i$ is independent of each other, and takes the periodic delivery sampling policy of sensor updates which is one of the most common approaches in practice~\cite{corneo2019age}. The sampling intervals of IoT devices are denoted as $ \mathrm{T} = \left \{ \tau_1,\tau_2,...,\tau_N \right \} $. The MEC server functions as a MEC server provider which is located in proximity to the IoT devices, and can be accessed by the IoT devices via the wireless channel. IoT devices that transmit computing tasks to the MEC server will be associated with clones at MEC server which execute computing tasks on behalf of IoT devices~\cite{mao2016dynamic}. Considering the constraint of IoT devices' computation capability, each IoT device can choose to process the status information locally with its own processor or to offload the sensed data to the MEC server or more powerful computing resources. After extracting the required information from the raw status data, the computing results are transmitted to the BS for future system control. To keep the freshness of the status information and guarantee the accurate control, the sensing, transmission and processing process needs to be executed repeatedly. Fig. \ref{fig1} shows the status offloading and processing procedure. 

\subsection{Sensing Model}
In this section, we describe the data sensing process of generating status updates for IoT devices. Let $t_i^{\text{unit}}$ be the sensing time for IoT device $i$ to perform a whole sensing task once. The status update processing task generated by IoT device $i$ can be represented as a tuple $U_{i}=\left\{d_{i},c_{i}\right\}$, where $d_{i}$ denotes the size of the sensory data generated in one sensing operation and $c_{i}$ denotes the necessary CPU cycles to finish the computing tasks. 

To evaluate the sensing quality of IoT devices, we utilize the probabilistic sensing model proposed in \cite{shakhov2017experiment}. When an IoT device executes a sensing task, the successful sensing possibility is denoted as 
\begin{equation}
    \varrho _i=e^{-\xi D^s_i},
\end{equation}
where $D^s_i$ is the distance between the IoT device and the status changing event and $\xi$ is a positive parameter to evaluate the quality of IoT device detection depending on the environmental condition. Considering one single data sensing operation will not satisfy the sensing successful possibility requirement, the IoT device may repeat the sensing operation to improve the sensing successful possibility. Let $S=\left \{s_1, s_2, ..., s_N  \right \} $ be the number of sensing operations executed by the IoT devices in a sensing operation. After finishing multiple sensing operations, the sensing successful possibility is denoted as 
\begin{equation}
    P_i\left ( s_i \right ) = 1-\left ( 1-\varrho_i \right ) ^{s_i}.\label{sen1}
\end{equation}
To ensure the sensing quality, the successful sensing possibility should be lower bounded. Let $p_{min}$ be the threshold for the successful sensing possibility of IoT devices. When IoT devices execute sensing operations, the successful sensing probability should satisfy:
\begin{equation}
    P_i\left ( s_i \right ) \ge p_{min},\ \forall i \in \mathcal{N}.
\end{equation}
Note that as the number of sensing operations increases, the IoT device achieves the higher sensing successful possibility. However, the multiple sensing operations will lead to a longer sensing time and larger sensing energy consumption. For IoT device $i$, the sensing time $T_i^{\text{ses}}$ is denoted as 
\begin{equation}
    T_i^{\text{ses}}\left ( s_i \right ) = t_i^{\text{unit}}s_i. 
\end{equation}
To process the generated status update in time, the sensing time should not exceed the sampling frequency. Otherwise, before the information of the status update is extracted, another new status update is generated by the IoT device, which make the information of the former status update outdated. Thus, the sensing time should satisfy the constraint:
\begin{equation}
    1\le s_i \le\left \lfloor\frac{\tau_i}{t^{\text{unit}}_i}   \right \rfloor.
\end{equation}
Let $e_{i}$ be the energy cost for IoT device $i$ to detect the status information per bit sensed data, the energy consumption of one single sensing process is 
\begin{equation}
    E_i^{\text{ses}}\left ( s_i \right ) = e_{i}d_is_i.
\end{equation}

It is worth noticing that the IoT devices will not figure out whether the status update is generating successfully from the raw sensory data. Further processing operations are required to verify the validity of the generated status update. If the sensing process is failed, the sensory data will be removed and the control unit will send the request of generating another new status update to the IoT device. If the IoT device has generated a new status update before receiving the request, the request will not be further tackled with. Otherwise, the IoT device will restart the sensing process immediately regardless of the sampling interval.

\subsection{Transmission Model}
When finishing the sensing procedure and generating a new status update, the IoT devices need to further process the status update to verify the validity of status update and extract the status information. Since some IoT devices are limited in computation capability, they need to transmit their status updates to the edge server for further processing. Let $\mathbf{x}  = \left \{ x_{1}, x_{2}, ...,x_{N} \right \} $ be the transmission policies for all the $N$ IoT devices, in which the elements can be expressed as follows.
\begin{equation}
x_i=\begin{cases}
 1, \text{ IoT device } i \text{ is transmitted to edge server.}\\
0,\text{ otherwise.} 
\end{cases}
\end{equation}
When the IoT device decides to transmit its status update to the edge server for processing, the transmission rate for the status update can be written as
\begin{equation}
      r_i(\mathbf{x})=B\log _2\left(1+\frac{g_{i,s} p_i}{\omega_0+ {\textstyle \sum_{m\in\mathcal{N},m\neq i}x_m}  g_{m,s}p_m}\right),
\end{equation}
\noindent where $B$ is the channel bandwidth allocated to the IoT device $i$, $g_i$ is the channel gain between device $i$ and the edge node, $p_i$ represents the device $i$'s transmission power, and $\omega_0$ represents the background interference power. Then, the transmission latency is calculated as 
\begin{equation}
    T_i^{\text{trans}} (\mathbf{x})= \frac{d_i}{r_i(\mathbf{x})}.
\end{equation}
Besides, the transmission energy consumption can be expressed by
\begin{equation}
    E_i^{\text{trans}} (\mathbf{x})= p_i \times \frac{d_i}{r_i(\mathbf{x})}.
\end{equation}
\subsection{Computation Model}
    In this subsection, we introduce the computation model of the status update. Dependent on the different computation offloading strategies taken by the IoT devices, the time and energy to execute the computation tasks of the status updates are different. If the IoT device choose to process the status update with its own local processor, the computation latency is denoted as
    \begin{equation}
        T_{i}^{\text{local}}=\frac{c_i}{f_i},
    \end{equation}
    \noindent where $f_i$ represents the CPU frequency of the local processor in IoT device $i$. Besides, processing the computation task locally brings extra energy consumption to the IoT device itself, which is calculated as 
    \begin{equation}
        E_{i}^{\text{local}}=c_i\delta.
    \end{equation}
    \noindent where $\delta$ is the energy consumption cost per CPU cycle.
    For the edge computing approach, the computation task is transmitted to the edge server for further processing and the computation latency is expressed by
    \begin{equation}
        T_{i}^{\text{edge}}=\frac{c_i}{f^e}.
    \end{equation}
    \noindent where $f^e$ denotes the CPU frequency of the edge server. 
    
    \section{Age of Information Analysis}
    \label{AoI}
    Note that the sensory information varies with time, so the freshness of status information has an essential impact on the accurate monitoring and controlling. We introduce the concept of AoI to evaluate the freshness of the status update generated by IoT devices. Let $A_i(t)$ be the AoI of the IoT device $i$ with $A_i(0) = 0$, and the AoI is defined as the time elapse from the last valid status update received by the control unit. One status update is valid to the control unit when the sensing process of the status update is success and the computation of the status update is finished. The AoI of the status update grows from the Beginning of the sensing, and keeps increasing until the next valid status update received by the control unit. Let $T_{i}^{j,\text{prcs}}$ be the processing time to conduct $j$ processing operations, and $T_{i}^{j,\text{prcs}}=j \times T_{i}^{1,\text{prcs}}$. Based on the different offloading strategies, the processing time of one single process operation is calculated as
    \begin{align} \label{sen2}
    T_{i}^{1,\text{prcs}}(s_i,\mathbf{x})&=T_i^{\text{ses}}\left ( s_i \right )+\left(T_i^{\text{trans}} (\mathbf{x})+T_i^{\text{edge}}\right)*x_{i}\notag\\
    &+T_i^{\text{local}}*(1-x_{i}).
    \end{align}
    However, the sensing time might be too short for IoT devices to successfully generate a valid status update with only one sensing operation. After finishing the computation execution, the control unit may find the uploaded status update fail to meet the requirement, i.e. the sensing processing is not successful. Under such scenario, the status update will be dismissed and a request will be sent to the IoT device to repeat the sensing process to generate another status update. 
    \begin{theorem} \label{theo1}
    For IoT device $i$, the average number of processing time for IoT devices to finish a processing task is $\frac{T_{i}^{1,\text{prcs}}\left (s_i,\mathbf{x}\right )}{P_i}$.
    \end{theorem}
    \emph{Proof:} Please see Appendix \ref{appendix1}.

     \begin{figure}
         \centering
         \includegraphics[width=0.95\linewidth]{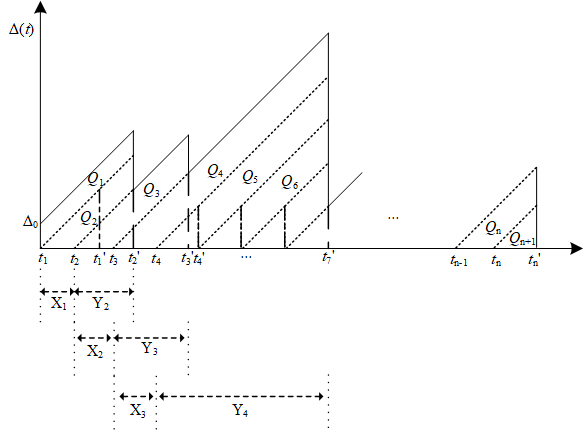}
         \caption{Example of AoI}
         \label{fig2}
     \end{figure}
     
Next, we derive the AoI of the status updates. Note that the information embedded in the status update makes sense after processing. The AoI at time $t$ of the $i$-th IoT device is denoted as 
\begin{equation}
    \Delta_i(t)=t-a_i(t),
\end{equation}
\noindent where $a_i(t)$ is the time when the latest status update generated by the IoT device $i$ is successfully sensed and processed. Without loss of generality, we assume the initial observing time is $t_1=0$, and the initial AoI is $\Delta_0$. As shown in the Fig. \ref{fig2}, the AoI grows linearly during the sensing and processing procedure and reset to a smaller value when a new status update is successfully accepted. Let $t_j$ be the time when the $j$-th status update generated and finish processing at the time $t_j'$. Let $Y_j$ denote the system time of the status update $j$, which is defined as 
\begin{equation}
    Y_j = t_u'-t_j, 
\end{equation}
\noindent where $t_u'$ is the time when the next status update is successfully sensed and executed. Specifically, when the status update $j$ is valid, $Y_j=t_j'$, i.e. the finishing time of the status update $j$. Besides, $X_j$ denotes the time between the generation of two continuous status update $j$ and $j+1$, which is given by
\begin{equation}
    X_j = t_{j+1}-t_{j}.
\end{equation}

Based on the definition above, the average AoI of the IoT device $i$ is denoted as 
\begin{equation}
    \overline{\Delta _i}=\frac{1}{T} \int_{0}^{T} \Delta_i\left ( t \right ) \mathrm{d}t.
\end{equation}
For the sake of simplicity, we consider the time interval with $T=t_n'$ which is displayed in Fig. \ref{fig2}. To calculate the average AoI, the area is divided into several geometric parts which is expressed by the concatenation of polygons $Q_j$. Hence, the average AoI is given by
\begin{equation} \label{eq1}
    \overline{\Delta _i}=\frac{1}{T} \sum_{j=1}^{n+1} Q_j.
\end{equation}
The area of the polygons are calculated differently. For $j=1$, $Q_1 =\Delta_0(X_1+Y_2)$. For $2 \le j \le n$, the area of the trapezoid $Q_j$ is calculated by the difference between two triangles, which is \begin{align}
    Q_j &= \frac{1}{2}(X_{j-1}+Y_j)^2-\frac{1}{2}Y_{j}^2 \notag\\
        &= \frac{1}{2}X_{j-1}^2+Y_{j}X_{j-1}.
\end{align}
Besides, the area of the $Q_{n+1}$ is the area of a triangle with a width of $Y_n$. Hence, the equation (\ref{eq1}) can be rewritten as 
\begin{align}
 \overline{\Delta _i} & = \lim_{T\to \infty} \frac{Q_1+Q_{n+1}+\sum_{j = 2}^{n}Q_j }{T} \notag\\
&= \lim_{T\to \infty} \left [\frac{Q_1+Q_{n+1}}{T} +\frac{n-1}{T} \frac{\sum_{j = 2}^{n}(\frac{1}{2}X_{j-1}^2+Y_{j}X_{j-1})}{n-1}\right ].
\end{align}
As $T$ becomes larger, i.e. $T\to \infty $, the value of $\frac{Q_1+Q_{n+1}}{T}$ can be ignored consequently and $\frac{n-1}{T}$ can be treated as the value of $\frac{1}{\mathbb{E}[X]}$. From the analysis above, we have
\begin{align}
    \overline{\Delta _i}&=\frac{\mathbb{E}[Q] }{\mathbb{E}[X] } =\frac{\frac{1}{2} \mathbb{E}[X^2]+ \mathbb{E}[XY]}{\mathbb{E}[X]}. 
\end{align}
Considering the sensing model mentioned above, the value of $\mathbb{E}[X]$ is dependent on the the sampling interval, which is given by $\mathbb{E}[X] = \tau_i$.

Besides, the value of $\mathbb{E}[Y]$ is identical to the average processing time of a successful status update $\mathbb{E}[T_i^{\text{prcs}}]$. Since $X_j$ is independent of $Y_j$, we derive the average AoI as
\begin{align}\label{aoi}
    \overline{\Delta _i}\left (s_i,\tau_i,\mathbf{x}\right )& = \frac{1}{2}\mathbb{E}[X]+\mathbb{E}[Y] \notag\\
&=\frac{1}{2}\tau_i +\frac{T_{i}^{1,\text{prcs}}\left (s_i,\mathbf{x}\right )}{P_i(s_i)}.
\end{align}
\section{Problem Formulation}\label{pf}
\subsection{Problem Formulation}
For a IoT device, the freshness of the generated status update plays a key role in accurate monitoring and controlling. Hence, the AoI of the status update should be well considered when evaluating the performance of the IoT device. Besides, energy consumption is another significant performance metric due to the physical constraint of IoT devices, and the sensing, transmission and computation operations all consume the energy during the IoT devices running procedure. The average value of IoT device $i$ is dependent on the energy consumption per time slot, which is given by
\begin{align}\label{energy}
    &\overline{E_i}\left (s_i,\tau_i,\mathbf{x}\right )\notag\\
    &=\lim_{T\to \infty} \frac{\sum_{j=1}^{n}[E_i^{\text{ses}}(s_i)+E_i^{\text{tran}}(\mathbf{x})*x_i+E_i^{\text{local}}*(1-x_i)]}{T}\notag\\
    &=\frac{E_i^{\text{ses}}(s_i)+E_i^{\text{tran}}(\mathbf{x})*x_i+E_i^{\text{local}}*(1-x_i)}{\mathbb{E}[X]}\notag\\
    &=\frac{E_i^{\text{ses}}(s_i)+E_i^{\text{tran}}(\mathbf{x})*x_i+E_i^{\text{local}}*(1-x_i)}{\tau_i}.
\end{align}
As discussed above, the overhead of the computation offloading problem includes both the average AoI and energy consumption, which can be formulated as
\begin{equation}
    C_i(s_i, \tau_i, \mathbf{x})= \mu_t\overline{\Delta _i}\left (s_i,\tau_i,\mathbf{x}\right )+\mu_e\overline{E_i}\left (s_i,\tau_i,\mathbf{x}\right ),
\end{equation}
\noindent where $\mu_t$ and $\mu_e$ are the weight parameters of the AoI and energy consumption respectively. For all the IoT devices, the optimization objective is to minimize the total overhead, which is expressed by
\begin{subequations}
\begin{align}
    &\mathcal{P}:\ \min_{S, \tau, \mathbf{x} } \sum_{i=1}^{N} C_i(s_i, \tau_i, \mathbf{x})\label{p1}\\
    &s.t.\ 1\le s_i \le\left \lfloor\frac{\tau_i}{t^{\text{unit}}_i}   \right \rfloor\ \forall i \in \mathcal{N},\label{c1}\\
    &\quad\quad x_i\in \left \{ 0,1 \right \},\ \forall i \in \mathcal{N} \label{c3}\\
    &\quad\quad \tau_i \ge \tau_{min},\ \forall i \in \mathcal{N} \label{c4}\\
    &\quad\quad P_i\left ( s_i \right ) \ge p_{min},\ \forall i \in \mathcal{N}\label{c6}\\
    &\quad\quad \sum_{i=1}^{N} x_id_i\le D_e.\ \forall i \in \mathcal{N} \label{c7}
\end{align}
\end{subequations}
\noindent where $D_e$ is the data threshold of the MEC server.
Constraints in (\ref{c1}) ensure the sensing time for a IoT device will not exceed the sampling interval. Constraint (\ref{c3}) guarantees the offloading decision for each IoT device is binary. Constraint (\ref{c4}) is the lower bound of the sampling interval for IoT devices. Constraints (\ref{c6}) is the lower bound of the successful sensing probability. Constraint (\ref{c7}) means the upper bound of the data size of the MEC server.
\subsection{Problem Decomposition}
Considering $\mathbf{s}$ and $\mathbf{x}$ are both discrete variables, the feasible set of Problem (\ref{p1}) is non-convex. Besides, the variables contains both continuous variables and discrete variables, which makes the optimization problem NP-hard \cite{sipser1996introduction}. In this part, we decompose the optimization problem into several subproblems: sampling interval optimization, sensing optimization and computation offloading optimization.

\paragraph*{Sampling Interval Optimization} Due to the constraint (\ref{c1}), the upper bound of the sensing time is dependent on the sampling interval of IoT devices. Hence, to obtain the optimal sensing time, the sampling period should be determined first. Note that the sampling interval has a great influence on the AoI and energy consumption of IoT devices. When the IoT devices generates status updates more frequently, i.e. the smaller $\tau_i$ for IoT device $i$, the AoI decreases accordingly. However, the energy consumption will increase greatly due to the frequent sampling action. In this subproblem, we study the optimal sampling interval for IoT devices to achieve the trade-off between the AoI and energy consumption, which is denoted as 
\begin{align}
\mathcal{P}_1:\ &\min_{\mathrm{T}} \sum_{i=1}^{N} C_i(\tau_i)\label{p2}\notag\\
    &s.t.\ (\ref{c4}).
\end{align}
\paragraph*{Sensing Time Optimization} Based on the result of the sampling interval optimization, the upper bound of sensing time is fixed. With the more sensing times, the sensing successful probability is greatly improved. However, the excessive sensing operation may lead to unnecessary sensing latency and extra sensing energy consumption. To determine the suitable sensing time for IoT devices, the problem $\mathcal{P}$ is rewritten as 
\begin{align}
\mathcal{P}_2:\ &\min_{S} \sum_{i=1}^{N} C_i(s_i)\label{p3}\notag\\
    &s.t.\ (\ref{c1}),(\ref{c6}).
\end{align}

%
\paragraph*{Computation Offloading Optimization} After solving $\mathcal{P}_1$ and $\mathcal{P}_2$, our goal is to find an optimal computation offloading policy for all the IoT devices to minimize the system overhead. The problem can be expressed as
\begin{align}
\mathcal{P}_3:\ &\min_{\mathbf{x}} \sum_{i=1}^{N} C_i(\mathbf{x})\label{p4}\notag\\
    &s.t.\ (\ref{c3}),(\ref{c7}).
\end{align}
Although the MEC server is equipped with powerful computation capability, the more IoT devices choose to transmit computing tasks to MEC server will cause severe interference which may lead to extra time consumption. Based on the observations, we aim to optimize the computation offloading strategies for IoT devices to minimize the system overhead.

\section{Joint Optimization of Sensing and Computation}\label{method}
In this section, we optimize the sampling interval, sensing time, and computation offloading optimization respectively by solving the subproblems proposed above. Then, we design an iterative algorithm to solve the problem $\mathcal{P}$ to minimize the system overhead jointly.
\subsection{Sampling Interval Optimization}
In this part, we solve the sampling interval optimization problem $\mathcal{P}_1$ mentioned in (\ref{p2}). Given the fixed sensing time and computation offloading policy, the value of $\frac{T_{i}^{1,\text{prcs}}}{P_i}$ remains unchanged. Therefore, for sake of simplicity, the expression of $\mathcal{P}_1$ can be rewritten as 
\begin{align}
\mathcal{P}_1:\ &\min_{\mathrm{T}} \sum_{i=1}^{N} C^{t}_i(\tau_i)\label{p42}\notag\\
&= \sum_{i=1}^{N}\bigg(\frac{\mu_t\tau_i}{2}+\notag \\
&\mu_e \frac{E_i^{\text{ses}}(s_i)+E_i^{\text{tran}}(\mathbf{x})*x_i+E_i^{\text{local}}*(1-x_i)}{\tau_i} \bigg) \notag\\
&s.t.\ (\ref{c4}).
\end{align}
 Since $\tau_i$ is the continuous variable, we calculate the the derivative directly to discuss the variation trend to address the optimization problem. Then, the derivative of $C^t_i(\tau_i)$ is calculated as
\begin{equation}
    \frac{\partial C^t_i(\tau_i)}{\partial\tau_i  } =\frac{\mu_t}{2} -\frac{\mu_eE_i^{\text{total} }}{\tau_i^2},
\end{equation}
\noindent where $E_i^{\text{total} }=E_i^{\text{ses}}(s_i)+E_i^{\text{tran}}(\mathbf{x})*x_i+E_i^{\text{local}}*(1-x_i)$ is a constant.
By solving $\frac{\partial C^t_i(\tau_i)}{\partial\tau_i  }=0$, we derive 
\begin{equation}
    \tau_i^*=\sqrt{\frac{2\mu_eE_i^{\text{total} }}{\mu_t}}.
\end{equation}
Due to the derivative of $C^t_i(\tau_i)$ are positive when $\tau_i>\tau_i^*$ and $C^t_i(\tau_i)$ is monotonic increasing, the optimal sample interval will be the lower bound if $\tau_i^*<\tau_{min}$. Hence, the optimal sampling interval is expressed as
\begin{equation}\label{sampleinterval}
     \tau_i^* = \left\{\begin{array}{l}
\sqrt{\frac{2\mu_eE_i^{\text{total} }}{\mu_t}}, \text{if }\sqrt{\frac{2\mu_eE_i^{\text{total} }}{\mu_t}}>\tau_{min},   \\
\tau_{min},\text{ otherwise.} 
\end{array}\right.
\end{equation}
\subsection{Sensing Time Optimization}
In this part, the subproblem (\ref{p3}) is considered to determine the optimal number of the sensing time. Note that the value of $T_i^{1,\text{prcs}}$ and $P_i$ increases with $s_i$ and the value of $C_i(s_i)$ increases with $T_i^{1,\text{prcs}}$ and decreases with $P_i$.  Besides, when the offloading policy is fixed, the cost of transmission and processing is determined. From the analysis above, the value of system overhead is rewritten with respect to the sensing time as 
\begin{align}
\mathcal{P}_2:\ &\min_{S} \sum_{i=1}^{N} C^{s}_i(s_i)\label{p42re}\notag\\
&= \sum_{i=1}^{N}\bigg(\mu_t\frac{T_i^{\text{ses}}(s_i)+T_i^{\text{ex} }}{P_i(s_i)}+\mu_e \frac{E_i^{\text{ses}}(s_i)+E_i^{\text{ex} }}{\tau_i} \bigg) \notag\\
&=\sum_{i=1}^{N}\left ( \mu_t\frac{t_i^{\text{unit}}s_i+T_i^{\text{ex}}}{1-\left ( 1-p_i \right ) ^{s_i}} + \mu_e\frac{e_{i}d_is_i+E_i^{\text{ex}}}{\tau_i} \right )\notag \\ 
&s.t.\ (\ref{c4}).
\end{align}
where $T_i^{\text{ex}}=\left(T_i^{\text{trans}} (\mathbf{x})+T_i^{\text{edge}}\right)*x_{i}+T_i^{\text{local}}*(1-x_{i})$ and $E_i^{\text{ex}}=E_i^{\text{tran}}(\mathbf{x})*x_i+E_i^{\text{local}}*(1-x_i)$ are constant when the sampling interval and computation offloading policy remain unchanged.

To minimize the system overhead, the variation trend of the objective function needs to be considered. To change the non-convex feasible set into a convex set, we relax the discrete variable $s_i$ into real value variable as $s_i \in \left [ 0,+\infty  \right ] $. It can be verified that the function of $C^{s}_i(s_i)$ is convex. Therefore, the value of $C^{s}_i(s_i)$ first decreases with $s_i$ and increases with the increment of $s_i$ and there is only one optimal solution for $s_i$. However, it is hard to achieve the optimal sensing time by directly solving $\frac{\partial C_i^s(s_i)}{\partial s_i} =0$. Considering the sensing time is discrete and has a upper bound, an enumerating algorithm is proposed to find the optimal sensing time, which is shown in Algorithm \ref{a1}. For each IoT device, the sensing time is initially set as $s_i=1$. The number of sensing operations keeps increasing until the system overhead is no longer decreasing. Considering the computation complexity of Algorithm \ref{a1} is dependent on the sampling interval which is a constant and the number of IoT devices. Let $\overline{\tau}$ be the average value of the sampling interval. The optimal sensing time $s_i^*$ can be derived with the complexity no more than $\mathcal{O}\left ( \overline{\tau} N \right ) $, 
\begin{algorithm}[t]
	\caption{Enumerating for Sensing Time Optimization}     
	 \label{a1}       
	\begin{algorithmic}[1] 
	\Require $\mathbf{x},\mathrm{T}$.  
    \Ensure Optimal sensing times $S$. 
    \For  {each IoT device $i$ $\in \mathcal{N}$}
    \State Initialization: $s_i=1;$
	\State Computing $C_i(s_i)\bigg|_{s_i=1}$;
	\While {$s_i \le\left \lfloor\frac{\tau_i}{t^{\text{unit}}_i}   \right \rfloor$}
	\If {$C_i(s_i+1)<C_i(s_i)$}
	\State $s_i = s_i +1$;
	\Else 
	\State break;
	\EndIf
	\EndWhile
	\EndFor
   \end{algorithmic} 
\end{algorithm} 
\subsection{Computation Offloading Optimization}
In this part, subproblem $\mathcal{P}_3$ is solved to determine the optimal computation offloading policy with the aim of minimizing the system cost. From the conclusion of \cite{chen2015efficient}, the computation offloading decision-making problem can be transformed into the maximum cardinality bin packing problem, which is NP-hard. Therefore, finding a central solution to the subproblem $\mathcal{P}_3$ is NP-hard. In view of the complexity of the offloading computation optimization problem, game theory is introduced to provide the decentralized way to conduct the computation offloading decision making. 

Before solving the offloading desicion making problem, the data size constraint of the MEC server need to be considered. Since the computation capacity of the MEC server is limited in practise, the number of data which can be processed at the same time is finite. When the data size of the MEC server exceed the threshold, the MEC server will not be able to serve the IoT devices any more. To meet the data constraint proposed in (\ref{c7}), we design a MEC server availability request mechanism, which can be utilized to request for the computing resource before offloading the computing task to the MEC server for processing. By virtue of the MEC server availability request mechanism, IoT devices can perform rational decision making without violating the data constraint of the MEC server.

Specifically, when an IoT device finishes the sensing operation and generates a status update, the IoT device first makes a preliminary offloading decision making based on the requirement of its computing task and the network condition. Then, the IoT device will send a computation capability request with the data size of the status update to the MEC server to apply for the equivalent computation capability of the MEC server. When the MEC server receives the request, the MEC server first summarize all its computation tasks to determine whether its computation capacity upper bound is exceeded and the amount of spare computation capacity to be allocated to IoT devices. If $\sum_{i=1}^{N} x_id_i\le D_e$, the MEC server will permit the computation offloading request and sent the computation capacity $f_e$ to the IoT device. Otherwise, the data size of computing tasks transmitted to the MEC server is beyond the data threshold of the MEC server. To reasonably allocate the computing resource of the MEC server, the MEC server will list the IoT devices that are being served by the MEC server and eliminate the computing task with the most value of the data size continuously until the data threshold $D_e$ is satisfied again. For those IoT devices are eliminated from the service list, the MEC server will send a message with the assigned value 0 of computation capacity to the deleted IoT devices. Consequently, the processing time for those IoT devices is infinite with 0 allocated computation capacity, and the IoT devices will choose the local processing instead. By virtue of the MEC server availability request mechanism, the MEC server unavailability is addressed and the subproblem $\mathcal{P}_3$ is transformed into the offloading decision making problem. Due to the data size of messages is relatively small, the communication overhead caused by the availability request mechanism can be ignored.

Then, we formulate the computation offloading decision making problem as a computation offloading game. Let $x_{-i}=\left \{   x_1,\dots ,x_{i-1},x_{i+1},\dots,x_{N}\right \}$ be the computation offloading policy of the other IoT devices except $i$. With the knowledge of the offloading strategies of other IoT devices, the IoT device $i$ perform the offloading decision making to minimize the system cost, i.e.
\begin{equation}
    \min_{x_i\in\mathbf{x} } C_i(x_i,x_{-i}),\forall i\in \mathcal{N}.\label{pgame} 
\end{equation}
The offloading decision making problem can be formulated as a strategic game $\Gamma = \left \{ \mathcal{N},\mathbf{x},C_i \right \} $, where the IoT device set $\mathcal{N}$ is the set of players, $\mathbf{x}$ is the set of strategies taken by players, and the system cost $C_i(x_i,x_{-i})$ is the objective function to be minimized. Then, we define the Nash equilibrium of the game $\Gamma$ as
\begin{definition}
A computation offloading strategy $\mathbf{x}^* = \left \{ {x}_1^*,...,{x}_N^* \right \} $ is a Nash equilibrium if no IoT devices can further reduce the system overhead by unilaterally changing its own computation offloading strategy, i.e.,
\begin{equation}
    C_i(x^*_i,x^*_{-i})\le C_i(x_i,x^*_{-i}),\forall x_i\in\left \{ 0,1 \right \} ,\forall i\in\mathcal{N}. 
\end{equation}
\end{definition}
For a multi-user computation offloading game, the Nash equilibrium guarantees that each IoT device at the Nash equilibrium will achieve a mutually satisfactory policy and have no incentive to deviate from its original strategy. The property is because if any IoT device is about to change its offloading policy, it should obtain lower system cost by updating offloading policy, which is contradictory to the definition the Nash equilibrium. Then, we define the best response for each IoT device:
\begin{definition}
For IoT device $i$, the strategy $x^*_i$ is the best response based on the policies of other users $x_{-i}$, if the system cost satisfies that
\begin{equation}
    C_i(x^*_i,x^*_{-i})\le C_i(x_i,x^*_{-i}),\forall x_i\in\left \{ 0,1 \right \}.
\end{equation}
\end{definition}
To achieve the Nash equilibrium, all the IoT devices tend to take the best response strategy. 
\begin{lemma}\label{lemma1}
An IoT device will achieve the lower system cost by offloading computing task to the MEC server for processing based on the offloading strategy $\mathbf{x}$, if the received interference meets ${\textstyle \sum_{m\in\mathcal{N},m\neq i}x_m}  g_{m,s}p_m \le L_i$, where $L_i$ is denoted as
\begin{align}
   L_i = \frac{g_{i,s}p_i}{2^{
\frac{\mu_td_i\tau_i+\mu_ep_id_iP_i}{B\left [  \mu_t\tau_i\left ( T_i^{\text{local}}-T_i^{\text{edge}} \right )+\mu_eP_iE_i^{local}\right ]  } }-1}-\omega _0.  \notag
\end{align}
\end{lemma}
\emph{Proof:} Please see the appendix \ref{appendix3}.

Accordingly, the best response of the IoT device $i$ can be expressed as
\begin{align}\label{bestresponse}
    x^*_i = \left\{\begin{array}{l}
1, \ \text{if }{ \sum_{m\in\mathcal{N}\setminus\left \{ i \right \} ,x_m  = 1}}  g_{m,s}p_m \le L_i, \\ 
0, \text{ otherwise} . 
\end{array}\right.
\end{align}
Based on Lemma \ref{lemma1}, the computation offloading strategy of IoT device $i$ is mainly dependent on its own received interference. To prove that the existence of the Nash equilibrium in our proposed computation offloading game, we introduce the concept of the potential game \cite{potentialgame}. 
\begin{definition}
A strategic game is called a potential game only if the variation of the utility function is proportional to the change of a certain function which is called potential function, i.e. there exists a potential function $\Phi(\mathbf{x}) $ satisfying that
\begin{align}
&C_i(x_i,x_{-i}) <C_i(x'_i,x_{-i}) \notag\\
&\text{iff}\ \Phi(x_i,x_{-i}) <\Phi (x'_i,x_{-i})
\end{align}
\noindent for each IoT device $i \in \mathcal{N}$, and any $x_i,x'_i\in\mathbf{x}$.
\end{definition}
\begin{theorem} \label{theo2}
The computation offloading decision making game is a potential game and always has at least one Nash equilibrium and possess the finite improvement property.
\end{theorem}
\emph{Proof:} Please see Appendix \ref{appendix2}.

According to Theorem \ref{theo2}, we know that the computation offloading decision making problem can achieve the Nash equilibrium after finite iterations. Next, we propose a decentralized computation offloading decision making algorithm in Algorithm \ref{a2} achieve the mutually satisfactory offloading strategy for IoT devices. 

\begin{algorithm}[t]
	\caption{Decentralized Computation Offloading Optimization Algorithm} 
	 \label{a2}       
	\begin{algorithmic}[1] 
	\Require $S,\mathrm{T}$.  
    \Ensure Optimal computation offloading strategy $\mathbf{x}^*$. 
    \State Initialize the computation offloading strategy that each IoT device chooses to process its task locally, i.e. $x_i=0,\ \forall i \in \mathcal{N}$;
    \Repeat \  {for each iteration slot $t$}    
    \State Initialize the update set $\kappa =\emptyset $.
    \For  {each IoT device $i$ $\in \mathcal{N}$}
       \State Compute the interference received by IoT device $i$ based on $x_{-i}$;
       \State Select the best response $x^*_i$ for $i$ according to (\ref{bestresponse});
       \If {$x_i$ $\ne$ $x^*_i$}
       \State Add IoT device $i$ into the update set $\kappa$ to compete for the updating opportunity;
       \Else
       \State Choose the original offloading strategy for next iteration slot $t+1$;
       \EndIf
    \EndFor
    \If{$\kappa=\emptyset$ }
    \State break;
    \EndIf
    \For{ each IoT device $i$ in the update set $\kappa$}
       \State Compute the improvement in the system cost by updating the offloading policy;
       \State Broadcast a request message with $\Delta C_i$ to contend for strategy update;
    \EndFor
    \For{ each IoT device $i$ in the update set $\kappa$}
    \If {$i$ possesses the most improvement in the system cost}
       \State Update the offloading strategy $x_i$ = $x^*_i$;
       \State Broadcast the decision update to all the other devices;
       \Else 
       \State Choose the original offloading strategy for next iteration slot $t+1$;
       \EndIf
    \EndFor
   \Until $\kappa=\emptyset$ for several consecutive slots
   \State\Return $\mathbf{x^*}$ 
   \end{algorithmic} 
\end{algorithm} 
To take the advantage of the finite improvement property of the potential game, we propose a decentralized computation offloading optimization algorithm to allow an IoT device to update its offloading strategy at one iteration. For each iteration slot, the update set is initialized as an empty set to record the IoT devices that have the incentive to update its offloading strategy. Based on the offloading strategies of other IoT devices $x_{-i}$, each IoT device computes its received interference by ${\textstyle \sum_{m\in\mathcal{N},m\neq i}x_m}  g_{m,s}p_m$. Then, the IoT devices will select the best response strategy according to (\ref{bestresponse}) and determine whether it needs to update its offloading strategy. If the best response is different from its current strategy, the device $i$ will be added to the update set to compete for the opportunity to update the offloading strategy. After all the IoT devices decide their best responses, the devices in the update set will evaluate the improvement range of updating the offloading policy by
\begin{equation}
    \Delta C_i =  C_i (x^*_i,x^*_{-i})- C_i (\overline{x^*_i} ,x^*_{-i}),
\end{equation}
\noindent where $\overline{x^*_i} = 1-x^*_i$ is the original strategy of the device $i$. To improve the convergence speed of the iteration, the IoT device with the most improvement will win the competition and update its offloading strategy. The other devices will sustain their original offloading strategy and wait for the next iteration to contend for the updating opportunity. The offloading strategy will be continuously iterated until no device tends to update its offloading strategy for several consecutive iterations, and the optimal offloading policy $\mathbf{x}^*$ is obtained. Since the most operations in Algorithm \ref{a2} are basic mathematical calculations, the computational complexity of one iteration is mainly dependent on the sort of the device with the most improvement. Since each device needs to perform the sorting operation, therefore the complexity of one iteration is $\mathcal{O}\left ( N\log_{}{N}  \right ) $. Assuming that $I$ iterations are required to achieve the Nash equilibrium, the complexity of Algorithm \ref{a2} is $\mathcal{O}\left ( IN\log_{}{N}  \right ) $.
\subsection{Algorithm Summary}
In this subsection, we summarize the multi-variable iteration system cost optimization algorithm (MISCO) for joint sensing and processing optimization to minimize the system overhead. To solve the overall optimization problem (\ref{p1}), we execute the iterations of the sensing, transmission and computation offloading optimization. First, we solve the optimal sampling period for each IoT device. Afterwards, based on the upper bound of the samping interval, we utilize the enumeration method to determine the optimal sensing time. Based on the result of the sampling and sensing optimization, a game-theoretic optimization algorithm is proposed to solve the optimal computation offloading strategy. Iterations of the sampling interval, sensing time and computation offloading optimization terminate when the disparity of the overall system cost $\hat{C}=\sum_{i=1}^{N} C_i $ between two consecutive iteration is below the threshold $\epsilon$. The details of the proposed algorithm is summarized in Algorithm \ref{a3}.

Based on the analysis above, the sampling interval set, the sensing time set and the computation offloading strategy are updated during the iteration process and the overall system cost keeps decreasing in each iteration. Considering the system cost has a lower bound and can only decrease finitely, the proposed multi-variable iterative optimization algorithm is convergent. Assuming $K$ iterations are requisite to meet the disparity threshold, the complexity of Algorithm \ref{a3} can be expressed as $\mathcal{O}\left ( KN+K\overline{\tau}N+KIN\log_{}{N}  \right ) $.
\begin{algorithm} 
	\caption{Multi-Variable Iterative System Cost Optimization Algorithm}     
	 \label{a3}       
	\begin{algorithmic}[1] 
	\Require Status update set $U$, distance set $D$, computation capacity $f_i$ and $f_e$, $\omega_0$, channel gain $g$, transmission power $p$, sensing time unit $t_i^{\text{unit}}$.  
    \Ensure Sampling interval $\mathrm{T}^*$, sensing time set $S^*$, optimal computation offloading strategy $\mathbf{x}^*$. 
    \State Set $r=0$ as the iteration slot. Initialize the sampling interval set $\mathbf{T}^0$, the sensing time set $S^0$ and the computation offloading strategy profile $\mathbf{x}^0$ randomly;
    \Repeat \  {for each iteration slot $r$}    
    \State Given the fixed $S^r$ and $\mathbf{x}^r$, solve the sampling interval $\mathbf{T}^{r+1}$ according to (\ref{sampleinterval});
    \State Given the $\mathbf{T}^{r+1}$ and $\mathbf{x}^r$, solve the sensing time $S^{r+1}$ using Algorithm \ref{a1};
    \State Given the $\mathbf{T}^{r+1}$ and $S^{r+1}$, solve the computation offloading strategy $\mathbf{x}^{r+1}$ using Algorithm \ref{a2};
    \State Compute the overall system cost $\hat{C}^{r+1}$ based on $\mathbf{T}^{r+1}$, $S^{r+1}$ and $\mathbf{x}^{r+1}$;
    \State $r = r+1$;
    \Until $|\hat{C}^{r}-\hat{C}^{r-1}|<\epsilon$
    \State\Return $\mathbf{x^*}$ 
    \end{algorithmic} 
\end{algorithm} 

\section{Simulation Results}\label{expresult}
In this section, we evaluate the performance of our proposed system overhead minimization algorithm by numerical results. We assume the coverage of the MEC server is a 50m$\times$50m area, and $N$ IoT devices are randomly distributed in the coverage area to execute the sensing tasks and generate the status updates. For sensing process, the time for executing a sensing task $t_i^{\text{unit}} = 0.2$ s, the sensing parameter $\epsilon = 0.08$ and the energy consumption for sensing data $e_i$ is $10^{-9}$ Joules/bit \cite{zhangsense}. The task size of the status update to be offloaded $d_i$ is 500 KB and the number of CPU cycles required to process the status update $c_i$ is 1000 Megacycles. For wireless communication, the channel bandwidth is set as $B = 100$ Mhz, the transmission power of the IoT device $p_i$ is 100 mW, and the background interference $\omega_0$ is -100 dBm \cite{chen2015efficient}. The channel gain of each IoT device is calculated as $g_{i,s} = v^{-o}$, where $v$ is the distance between the IoT device and the MEC server and $o$ is path loss coefficient set as 4. The computation capability for the IoT device $f_i$ is in a range of [0.8,1.0] Ghz, and the computation capability of the MEC server is 20Ghz. The processing power for IoT device to execute the computing tasks locally per CPU cycle is $\delta = 10^{-11}(f_i)^2$ \cite{chen2015efficient}.

\begin{figure}[t]
    \centering
    \includegraphics[width = 8cm]{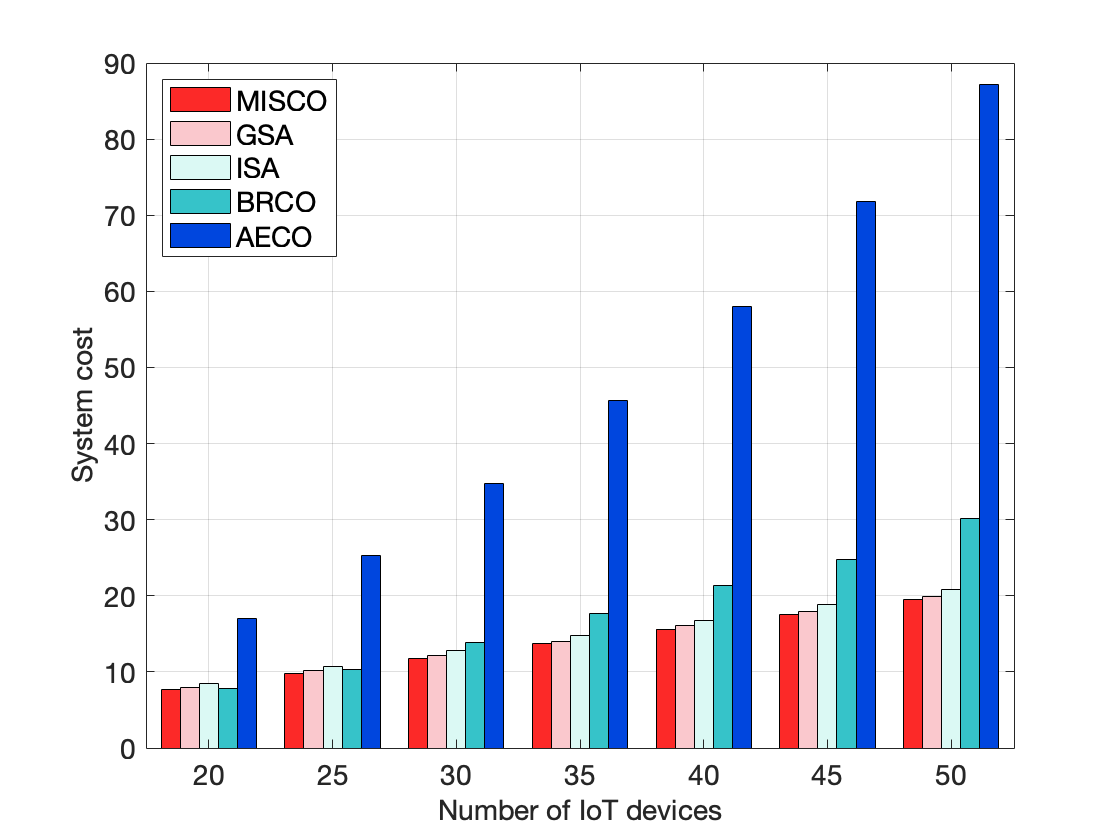}
    \caption{System cost vs. number of IoT devices.}
    \label{fig3}
\end{figure}

To evaluate the effectiveness of our proposed algorithm, we introduce four comparative algorithm as benchmark:
\begin{itemize}
\item Greedy Sensing Algorithm (GSA): The sampling interval is decided according to (\ref{sampleinterval}). The IoT devices will execute the least sensing operations to meet the sensing successful probability to achieve the least sensing latency. Then, the offloading decision-making is make by Algorithm \ref{a2}.
\item  Instant Sampling Algorithm (ISA): Each time when the previous status update finishes processing, the IoT device will conduct another sampling process to generate a new status update, which is similar to the zero-wait policy in \cite{barakatis}. To minimize the time cost, the IoT device will generate the status update instantly without waiting time. The sensing time decision is made by Algorithm \ref{a1} and the computation offloading strategy is made by Algorithm \ref{a2}.
\item Best Response Computation Offloading (BRCO) \cite{wanggame}: The sampling interval is decided by (\ref{sampleinterval}), and the sensing time is obtained from Algorithm \ref{a1}. Each device chooses the best-response strategy based on the computing cost of the two processing ways and the offloading probabilities of other devices in the previous stage. Then, each device chooses its offloading strategy according to the converged offloading possibility.
\item All Edge Computation Offload (AECO): The sampling interval and sensing time are determined in the same way as our proposed optimization algorithm. Then, all the computing tasks are offloaded to the MEC server for processing.
\end{itemize}
We first evaluate the system cost of our proposed method. Fig. \ref{fig3} shows the system cost of different methods with the different numbers of IoT devices. Compared with the other benchmarks, our proposed MISCO achieve the lowest system cost as the number of IoT devices increases. When the number of IoT devices is relatively small, there is not much difference between AECO and the other four comparative methods due to the small interference caused by IoT devices. Compared with the other two computation offloading optimization methods BRCO and AECO, our proposed method has a better performance in system cost as the number of IoT devices increases. With the larger number of IoT devices, the bandwidth for IoT devices to execute the task transmission is insufficient and the transmission cost improves greatly. Therefore, our computation offloading method perform a more rational offloading decision making.

\begin{figure}[t]
    \centering
    \includegraphics[width = 8cm]{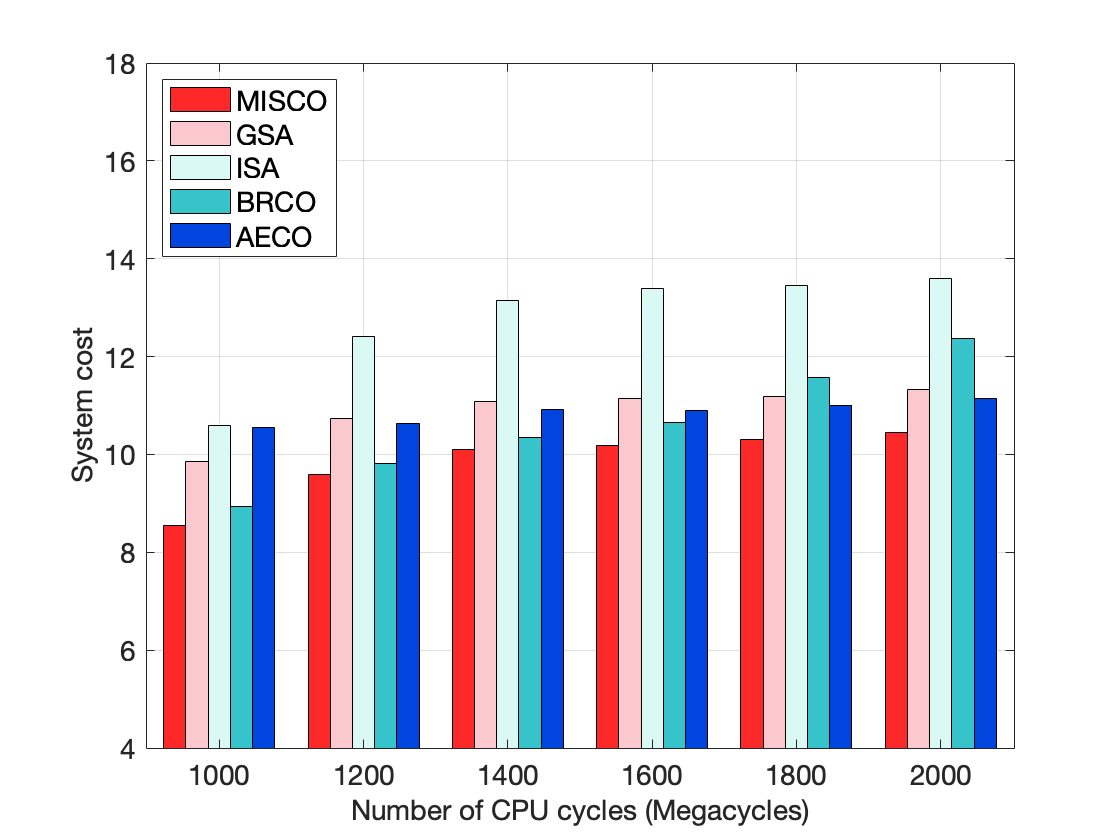}
    \caption{System cost vs. number of CPU processing cycles.}
    \label{fig4}
\end{figure}

Fig. \ref{fig4} shows the system cost with the different numbers of the CPU cycles required to process the status update. The numerical result displays the system cost of MISCO is lower than other benchmarks with more computation load of the status update. The system cost of ISA rises greatly because the longer process time of the status update make the status sampling out of date, while the increment of system cost of sensing optimization is relatively stable as the number of CPU cycles increases. With the improvement of the CPU cycle, the performance of MISCO will be closed to the AECO. Due to the hard computation load, all the IoT devices will choose to offload their tasks. 

\begin{figure}[t]
    \centering
    \includegraphics[width = 8cm]{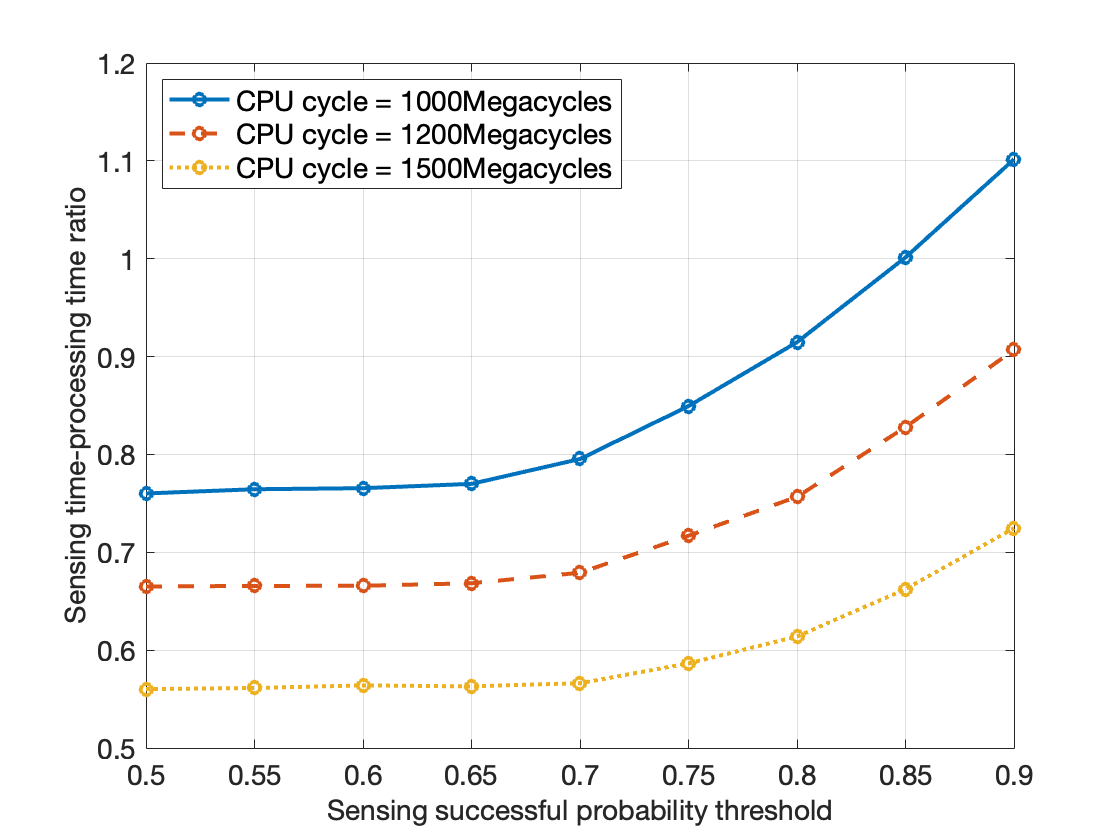}
    \caption{Sensing time-processing time ratio vs. successful sensing probability thresholds.}
    \label{fig5}
\end{figure}

Fig. \ref{fig5} depicts the sensing time-processing time ratio with the different successful sensing probability under various computation loads. Note that the processing time here consists of the transmission time and the task processing time. When the sensing successful probability threshold is relatively small, the sensing time-processing time ratio remains a stable level. Considering the threshold is easy to meet, the sensing time is determined by the sensing optimization method. With the requirement of the sensing successful probability threshold increases, more sensing operations need to be performed to satisfy the successful probability threshold, which causes more sensing time for IoT devices. Specifically, when the CPU cycle is 1000 Megacycles and the sensing successful probability threshold is more than 0.65, the sensing successful probability threshold has a great influence on the sensing-processing ratio. That is, the processing time dominates the ratio the sensing time-processing time ratio when the successful probability requirement is not strict and then the sensing time dominates with the high successful probability requirement. As the processing CPU cycle increases, the sensing time dominates at a higher level of successful probability due to the more processing time for status update. When the the CPU cycle is 1500 Megacycles, the sensing-processing ratio begin to increase when the  sensing successful probability threshold is more than 0.7.

\begin{figure}[t]
    \centering
    \includegraphics[width = 8cm]{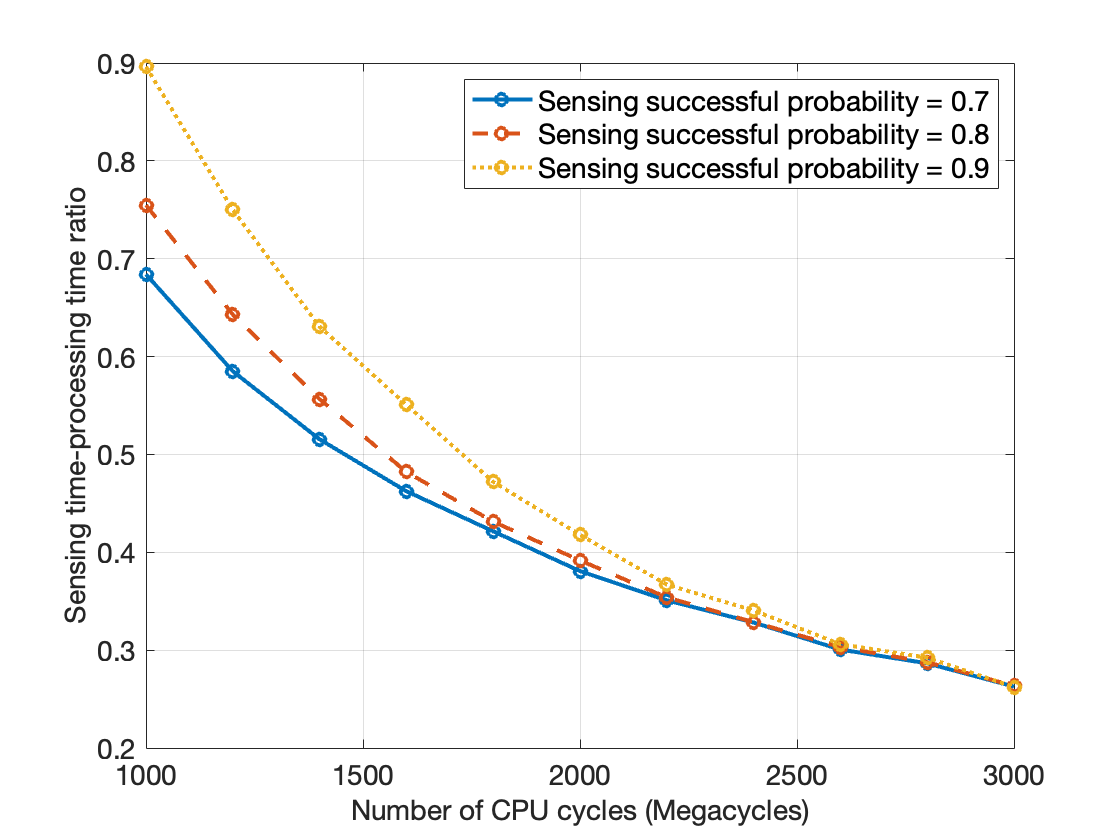}
    \caption{Sensing time-processing time ratio vs. number of CPU processing cycles.}
    \label{fig6}
\end{figure}

The simulation result of the sensing time-processing time ratio with the different CPU cycles is shown in Fig. \ref{fig6}. When the computation load is small, the sensing time dominates the sensing-processing ratio. With the higher sensing successful probability threshold, the sensing time accounts for a higher proportion of the total time cost. As the number of CPU cycles increases, the 
sensing time as a percentage of total time decreases sharply. In the case of large number of CPU cycles, the sensing successful probability has little effect on the ratio of sensing time to processing time. The sensing-processing ratio remains at the same level for different successful probability thresholds, and it can be seen that processing time accounts for the major part of the total time at high computation load.

\begin{figure}[t]
    \centering
    \includegraphics[width = 8cm]{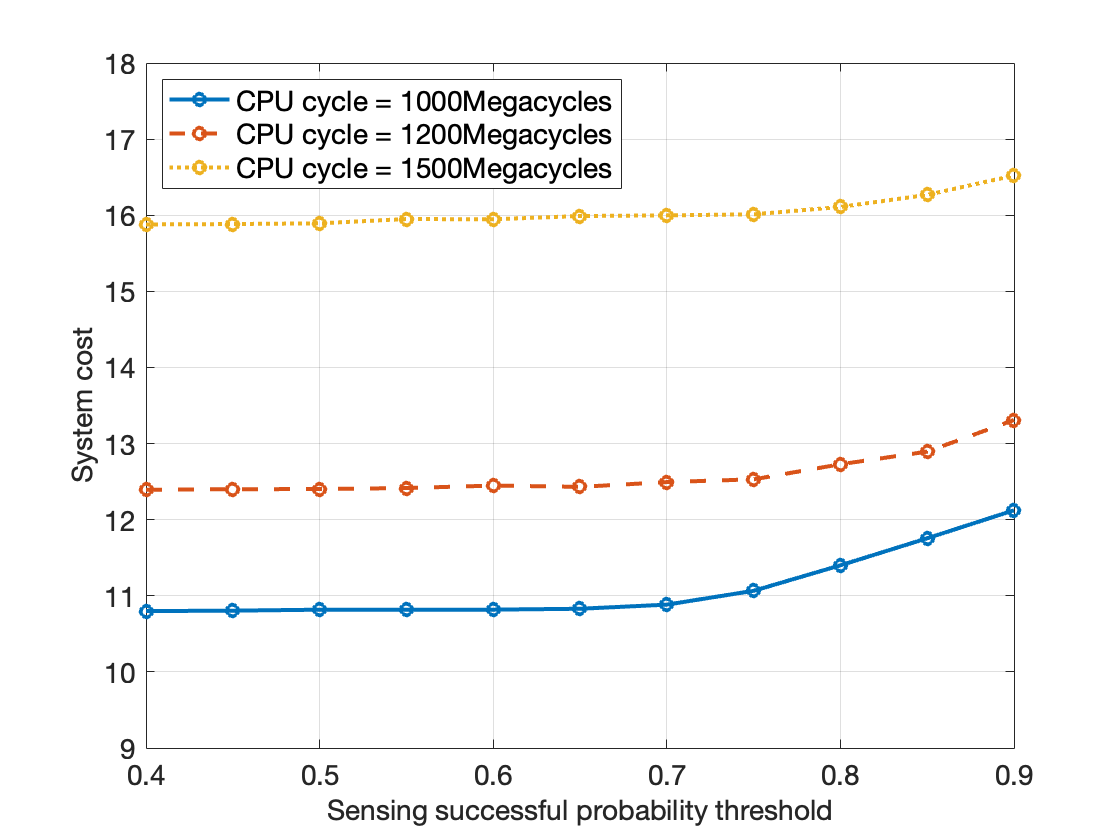}
    \caption{System cost vs. successful sensing probability thresholds.}
    \label{fig7}
\end{figure}

Fig. \ref{fig7} shows the impact of the different sensing successful probability threshold on the system cost. When the threshold value is at a low level, the system consumption will not be limited by the threshold and can be optimized to get the minimal system overhead directly by our proposed optimization algorithm. Therefore, the system cost will be maintained at a stable level. After the threshold value exceeds 0.7, the optimization method will be limited to achieve the optimal system overhead in order to meet the sensing successful rate requirement. When the CPU cycles for the task execution is small, the more obvious is the influence by the sensing successful rate threshold and the system cost rises more obviously. Therefore, the optimal threshold should be set to about 0.7 to ensure the sensing quality as well as the value of the system overhead.

\begin{figure}[t]
    \centering
    \includegraphics[width = 8cm]{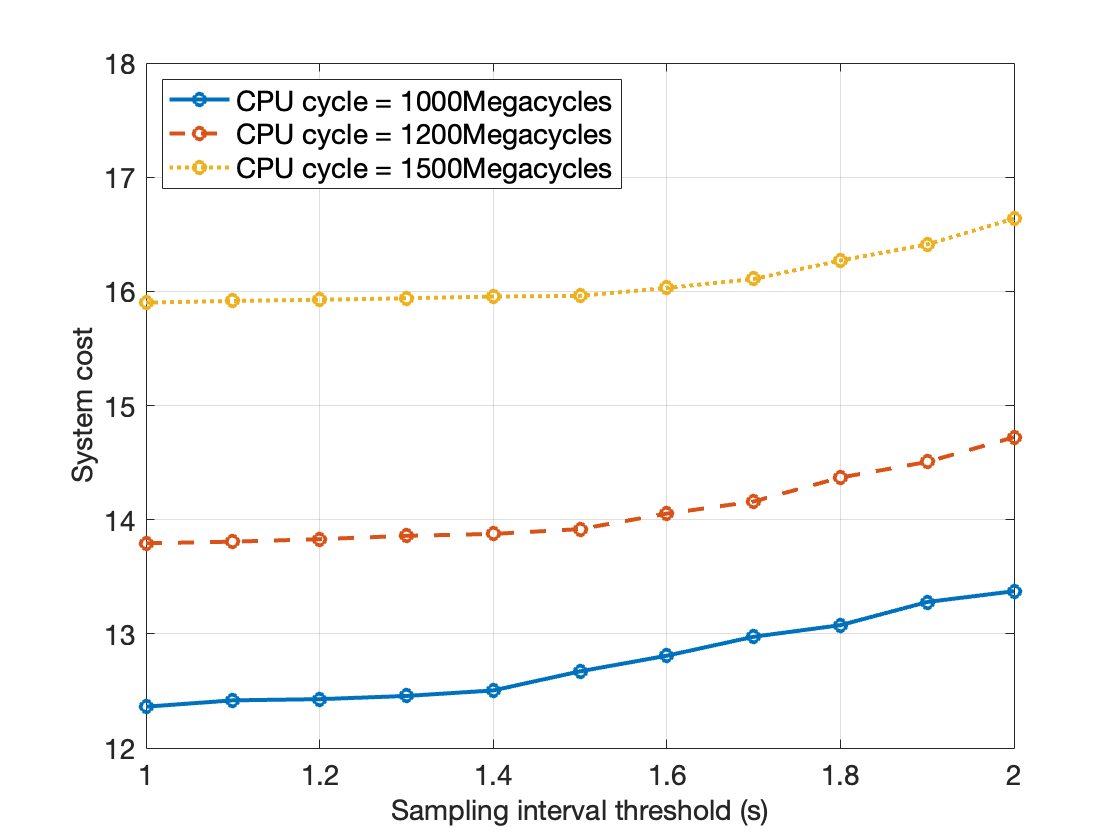}
    \caption{System cost vs. sampling interval thresholds.}
    \label{fig8}
\end{figure}

In Fig. \ref{fig8}, the numerical result shows the system cost of our proposed MISCO with different sampling intervals. With the high sampling frequency limit, the system overheads for different task computations load are sustained at a stable value. When the sampling interval is greater than 1.4 s, the system overhead all starts to gradually increase. This means that the sampling interval limit at this point is greater than the optimal sampling interval solved by our proposed algorithm, and the excessively long sampling interval leads to an increase in the AoI, resulting in an increase in system overhead. To reduce the energy consumption of generating status updates and the AoI of status update, the sampling interval threshold should be set lower than 1.4 s.

\begin{figure}[t]
    \centering
    \includegraphics[width = 8cm]{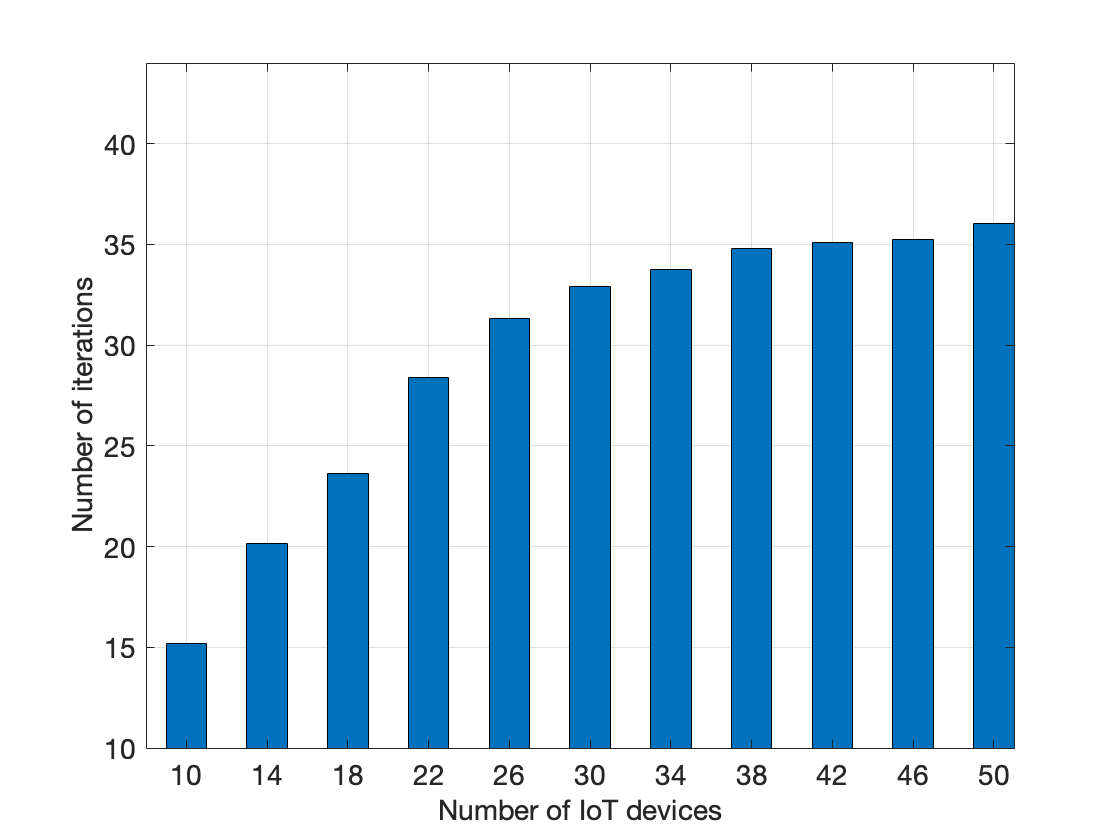}
    \caption{Number of iterations vs. Number of IoT devices.}
    \label{fig9}
\end{figure}

Fig. \ref{fig9} depicts the number of iterations of our proposed method with the different IoT devices numbers. Here the number of iteration contains the iteration number of Algorithm \ref{a2} to achieve the Nash equilibrium and the iteration number of Algorithm \ref{a3} to converge. The iteration number increases with the increasing of the IoT device number, which illustrates the convergence and scalability of our proposed MISCO. When the number of IoT devices is getting larger, the network resources are insufficient for the IoT devices. Therefore, all the IoT devices tend to process its computing tasks locally which makes the number of iterations reach an upper bound due to all devices stop iterating after choosing local execution.

\section{Conclusion}\label{conclusion}
In this paper, we first formulate the joint sensing and processing optimization problem to minimize the system overhead including the information freshness of the status updates and the energy consumption of IoT devices. The optimization problem is decomposed into three subproblems to optimize the sampling, sensing and computation offloading respectively. The sampling and sensing optimization problem are solved by extremum principles and game-theoretic method is utilized to perform the computation offloading decision-making. Afterwards, the multi-variable iterative optimization algorithm is proposed to minimize the system cost jointly. Numerical results depicts that the system cost achieved by our proposed method is lower than other comparative methods and the dominance of sensing and processing under different scenarios. Besides, the impact of the sensing probability and sampling interval thresholds are analyzed in the simulation.
\appendices
\section{Proof of Theorem \ref{theo1}} \label{appendix1} 
The probability that IoT device $i$ requires $j$ sensing operations to generate a valid status update is $P_i\left ( s_i \right )(1-P_i\left ( s_i \right ))^{j-1}$. Thus the expectation of execution time can be calculated as
\begin{align}
\mathbb{E} [T^{\text{prcs}}_i]
&= \lim_{j \to \infty } P_i\left ( s_i \right )*T_i^{1,\text{prcs}}(s_i,\mathbf{x})\notag\\
&+P_i\left ( s_i \right )*(1-P_i\left ( s_i \right ))T_i^{2,\text{prcs}}(s_i,\mathbf{x})\notag\\
&+ P_i\left ( s_i \right )*(1-P_i\left ( s_i \right ))^2T_i^{3,\text{prcs}}(s_i,\mathbf{x})+...\notag\\
&+P_i\left ( s_i \right )*(1-P_i\left ( s_i \right ))^{j-1}T_i^{j,\text{prcs}}(s_i,\mathbf{x})\notag\\
&=\lim_{j \to \infty } P_i\left ( s_i \right )*T_i^{1,\text{prcs}}(s_i,\mathbf{x})\sum_{n = 1}^{j} n(1-P_i\left ( s_i \right ))^{n-1}.
\end{align}
Given the value of $P_i\left ( s_i \right )$ is in the range of (0,1), let $1-P_i\left ( s_i \right )$ be a single variable $\rho$ and $\sum_{n = 1}^{\infty } n(1-P_i\left ( s_i \right ))^{n-1}$ can be further calculated as 
    \begin{align} \label{sumres}
\sum_{n = 1}^{\infty } n(1-P_i\left ( s_i \right ))^{n-1} & = \sum_{n = 1}^{\infty } {[(1-P_i\left ( s_i \right ))^n]}' \notag\\
&=\sum_{n = 1}^{\infty }{(\rho ^n)}' = {(\frac{\rho }{1-\rho }) }'\notag\\
&=\frac{1 }{(1-\rho)^2 }=\frac{1}{P_i\left ( s_i \right )^2}.
\end{align}
By substituting (\ref{sumres}) into the expectation of execution time, the average number of processing time for a successful status update is calculated as
     \begin{equation}
         \mathbb{E} [T^{\text{prcs}}_i]=\frac{T_{i}^{1,\text{prcs}}(s_i,\mathbf{x})}{P_i\left ( s_i \right )}.
     \end{equation}
\section{Proof of Lemma 1}\label{appendix3}
According to (\ref{sen2}), (\ref{aoi}) and (\ref{energy}), the lower system cost of transmitting computing task to the MEC server for processing is equivalent to
\begin{align}
    \mu_t\frac{T_i^{\text{local}}}{P_i} +\mu_e\frac{E_i^{\text{local}}}{\tau_i} \ge \mu_t\frac{T_i^{\text{edge}}+T_i^{\text{tran}}(\mathbf{x})}{P_i} +\mu_e\frac{E_i^{\text{trans}}(\mathbf{x})}{\tau_i} \notag
\end{align}
That is
\begin{align}
    r_i(\mathbf{x})\ge\frac{\mu_td_i\tau_i+\mu_ep_id_iP_i}{\mu_t\tau_i\left ( T_i^{\text{local}}-T_i^{\text{edge}} \right )+\mu_eP_iE_i^{\text{local}} } \notag
\end{align}
Then, we can derive the threshold of the interference that achieves the lower system cost by offloading tasks to the MEC server for execution
\begin{align}
\sum_{m\ne i}  x_mg_{m,s}p_m \le\frac{g_{i,s}p_i}{2^{
\frac{\mu_td_i\tau_i+\mu_ep_id_iP_i}{B\left [  \mu_t\tau_i\left ( T_i^{\text{local}}-T_i^{\text{edge}} \right )+\mu_eP_iE_i^{local}\right ]  } }-1}-\omega _0.  \notag \notag
\end{align}

\section{Proof of Theorem \ref{theo2}} \label{appendix2}
To prove the computation offloading decision making game is a potential game, we define the potential function as 
\begin{align}
\Phi(\mathbf{x}) & = \frac{1}{2} \sum_{i  = 1}^{N} \sum_{m\ne i} g_{i,s}p_ix_ig_{m,s}p_mx_m \notag\\
&+\sum_{i  = 1}^{N} g_{i,s}p_iL_i(1-x_i).
\end{align}
We consider an IoT device chooses to update its offloading policy with a lower system overhead, i.e. $C_i(x_i,x_{-n}) <C_i(x'_i,x_{-n})$. From the definition of the potential game, the decrease of the system cost will lead to the decrease of the potential function. If the original offloading policy is to process the task locally, i.e. $x'_i =0,\ x_i=1$, we derive $C_i(1,x_{-n}) <C_i(0,x_{-n})$, and ${ \sum_{m\in\mathcal{N}\setminus\left \{ i \right \} ,x_m=1}}  g_{m,s}p_m \le L_i$ is meet. Then, we compute the change of the potential function by updating the offloading policy:
\begin{align}
&\Phi(1,x_{-i}) -\Phi(0,x_{-i})\notag\\
&=\frac{1}{2}\sum_{j\ne i}\sum_{m\ne i,m\ne j} g_{j,s}p_jx_jg_{m,s}p_mx_m+\frac{1}{2}g_{i,s}p_i\sum_{j\ne i}g_{j,s}p_jx_j \notag \\
&+\frac{1}{2}g_{i,s}p_i\sum_{m\ne i}g_{m,s}p_mx_m +\sum_{j \ne i} g_{j,s}p_jL_j(1-x_j)\notag \\
&-\frac{1}{2}\sum_{j\ne i}\sum_{m\ne i,m\ne j} g_{j,s}p_jx_jg_{m,s}p_mx_m -\sum_{j \ne i} g_{j,s}p_jL_j(1-x_j)\notag\\
&-g_{i,s}p_iL_i\notag \\
&=g_{i,s}p_i\sum_{m\ne i}g_{m,s}p_mx_m -g_{i,s}p_iL_i<0.
\end{align}
For $x'_i =1, x_i=0$, the result is similar to the argument above. According to the definition of the potential game, we conclude that the computation offloading decision making problem is a potential game. 

\section*{Acknowledgment}

The authors would like to thank...

\ifCLASSOPTIONcaptionsoff
  \newpage
\fi




%

\end{document}